\RequirePackage{fix-cm}
\documentclass[]{interact}

\usepackage{hyperref}
\usepackage{amssymb}
\usepackage{amsmath}
\usepackage{url}
\usepackage{tabularx}
\usepackage{longtable}
\usepackage{epsfig}
\usepackage{cite}
\usepackage{setspace}
\usepackage{xtab}
\usepackage{graphicx}
\usepackage{epstopdf}
\usepackage{graphicx}
\usepackage{color}
\usepackage[numbers]{natbib}
\usepackage{subfig}
\usepackage{algorithm}
\usepackage{algorithmic} 
\usepackage{longtable} 
\usepackage[table,xcdraw]{xcolor}
\usepackage{pgfplots}

\usepackage{tikz}
\usetikzlibrary{shapes.geometric,arrows}\usepackage{pgfplots}

\theoremstyle{plain}

\theoremstyle{definition}

\theoremstyle{remark}

\begin{document}


\title{Improved Algoritms in Parallel Evaluation of Large Cryptographic S-Box}


\author{
\name{Behrooz Khadem* \textsuperscript{a}\thanks{} and Reza Ghasemi\textsuperscript{b}}
}

\maketitle

\begin{abstract}
Nowadays computational complexity of fast walsh hadamard transform and non-linearity for Boolean functions and large substitution boxes is a major challenge of modern cryptography research on strengthening encryption schemes against linear and differential attacks. Time and memory complexities of the best existing algorithm for computing fast walsh hadamard transform and non linearity for $\mathrm{n\times m}$ substitution boxes $(n\geq16, m\geq16)$ is ${O(\mathrm{2}}^{\mathrm{n+m}})$. This paper proposes three new acceleration methods that improve the computation time for parallelized walsh matrix up to 39 folds and the computation time for non linearity degree up to 563 folds, defining and accessing walsh matrix transpose, and incorporating an important part of computation process of non linearity in the computation algorithm of walsh matrix. The validity of the proposed algorithms is verified by means of simulation and experimentation and the overall analysis of resource consumption of proposed algorithms was compared with previous ones.
\end{abstract}


\section{Introduction}
\label{intro}
Boolean functions and substitution boxes (S-Boxes) are responsible for creating Shanon's entropy \cite{1A} in Stream and Block ciphers \cite{1B}. The fast walsh-hadamard transform (FWHT) is a method for simplifying presentation and computation of some encryption features of boolean functions as well as S-Boxes (such as nonlinearity and the degree of correlation immunity). The FWHT time and memory complexities of an $\mathrm{n\times m}$ S-Box is ${O(n2}^n2^m)$ and ${O(2}^n2^m)$, respectively. Moreover, computation of nonlinearity for S-Boxes is indispensable and inevitable given the effect of the resistance of S-Boxes to linear attacks \cite{2}. But the time and memory complexity of computation of nonlinearity in an $\mathrm{n\times m}$ S-Box is ${O(2}^n2^m)$. Therefore, acceleration and parallelization of FWHT computation and nonlinearity computation can have significant impacts on reducing computational complexity of this feature.

Given the exponential complexity of computing and memory in calculating nonlinearity and FWHT in a S-Box, the challenge of large scale S-Boxes (larger than $\mathrm{16\times 16}$) is far greater so that this challenge is also still one of the key open issues in cryptography \cite{3}. Based on recent researches, evaluating large S-Boxes requires a great amount of time and memory, so that the evaluation operation halted while implementing with MATLAB  and ordinary computers \cite{4}.

The reasons for using large S-Boxes include avoiding advanced side channel attacks and challenges created by quantum machines in post-quantum cryptography \cite{5A}. While these challenges act as practical threats to small S-Boxes, they are not virtually effective against large ones. Therefore, the use of large S-Boxes in block ciphers applied in conventional communication protocols (such as Kasumi in GSM) has recently drew attention of scholars \cite{5B}. It has also been shown that the large size of these S-Boxes can improve some of their encryption features, such as increasing their algebraic degree, increasing their resistance to differential and linear attacks \cite{6}, improving avalanche and strict avalanche feature \cite{7}, and providing bit independence \cite{8}. 

For large S-Boxes, walsh matrix computation and memory access requires a great amount time and memory complexities. Thus, this paper aims to improve parallelization methods and modify memory allocation and access as well as incorporating a part of nonlinearity computation algorithm (al.) into the walsh matrix computation al. for efficient reduction of time and memory consumed in calculations.

The rest of this paper is organized as follows: Section \ref{rb} introduces the research background. In Section \ref{bca}, basic required concepts and al.s are defined. Section \ref{dpc} discusses the main problem of the research and its secondary objectives. The proposed acceleration methods are introduced in Section \ref{ps}. Section \ref{ree} comprises comparison of analytical and experimental results obtained in experiments conducted before and after the acceleration. Finally, Section 7 provides an overview of the overall results and suggestions for further work.  

\section{Research Background} \label{rb}

Osusania et al. (1994) were first to calculate the FWHT for boolean functions by parallelization using ordinary computers and a supercomputer. However, they did not introduce any al.s and merely recorded the results \cite{9}. They succeeded in accelerating computations only by six times and explained the reason to be lack of an operating system capable of maximizing synchronization of the threads. 

Computation of nonlinearity of S-Boxes using FWHT was first carried out by Youssef et al. (1995) who obtained an estimation of nonlinearity \cite{10}. They presented an upper limit for nonlinearity using Linear Approximation Table (LAT) for S-Boxes, but they did not discuss computational and memory complexities of the method proposed in their paper.

Andrade et al. (2014) proposed a method for parallelizing of FWHT computation that utilized pipeline and hierarchical memory in Graphic Processing Units (GPUs) and thereby, they minimized memory bank constraints on GPUs \cite{11}. The authors noted in their paper that they have computed a maximum of 69,521 FWHT per time unit (milliseconds) for a boolean function with 128-bit input. They made no mention of extending their method to S-Boxes or even boolean functions larger than 265-bit inputs, nor did they describe their parallelization method.

Picek et al. (2014) presented the SET software for evaluating boolean functions and S-Boxes \cite{12}. Two main goals were considered in the design of this software: evaluating a wider set of features than similar tools and obtaining a S-Box with desired features. The authors acknowledged that in their research they did not observe the feature of bit independence and that their proposed method consumed a great deal of time and memory for large S-Boxes of 16 $\mathrm{\times}$ 16. They suggested examination of ways to improve their method for large S-Boxes.

Amaral et al. (2016) implemented a software tool to evaluate the cryptographic features of S-Boxes, including nonlinearity and FWHT, using different computational techniques \cite{4}. They suggested that it is easier to implement the tool with MATLAB for small-size S-Boxes while for larger ones, implementation in C ++ would be better because of efficient memory usage. They also concluded that implementation was not suitable for large S-Boxes in terms of time and memory and suggested use of parallelization for solving this problem.

Bikov et al. (2018) parallelized the walsh spectrum of boolean functions by using fast walsh al. in alternative ways, running them on GPU, and finally, comparing the results \cite{13}. Various parallelization of fast walsh al. in Bikov's paper have different techniques in terms of complexity of implementation, resources used, and optimization strategy. In his paper, Bikov computed the walsh spectrum only for one boolean function and S-Boxes were not mentioned. 

As it can be seen in the previous literature, two important challenges has remained. First, none of the previous studies has conducted parallelization of FWHT for S-Boxes (as vectorial boolean functions). Second, no evaluation have been conducted on nonlinearity of large S-Boxes so far. This paper responds to these two challenges by innovatively implementing a hybrid improvement method involving parallelization, memory management, and improvement of FWHT computation al. and its application in nonlinearity estimation of large S-Boxes. 

\section{Basic Concepts and Algorithms} \label{bca}

In parallel al.s, there is no dependency between commands or parts and different commands in a parallel al. can be executed concurrently (in one time unit). Parallelization is a feature by which those al. commands (which can be parallelized) are executed concurrently and in parallel by multiple processors or multiple processing cores.

In parallel al.s, the implementation of each parallel part should map to a thread. These threads are run concurrently in the CPU. The operating system can put threads in the sleep mode or even remove them from RAM and wake them up upon their own request or whenever it deems appropriate. The displacement between random access and storing memory is called swap operation. 


A S-Box is a generalization of a boolean function to a vectorial boolean function of $s=(f_1,f_2,\dots ,f_m)$ with a multi-bit output ($f_i$ is the function of the i-th component). The walsh-hadamard matrix $S:F^n_2\to F^m_2$ is the S-Box S as Equation \eqref{wh} \cite{15}.
\noindent 

\begin{equation} \label{wh}
\begin{split}
w_s(u,v) = 
\sum^{2^n-1}_{x=0}{{(-1)}^{<v,S\left(x\right)>+<u,v>}} \ \ \ \ 
\left(0\le u\le 2^n-1\right),\left(0\le v\le 2^m-1\right)
\end{split}
\end{equation}

Here the fast walsh al. presented in \cite{16} is used to compute the walsh matrix.  Al.~\ref{algorithm1} represents the fast walsh al. for computing the walsh matrix for a S-Box.

\begin{algorithm}
      \small
	\caption{Pseudcode of FWHT}
	\label{algorithm1}
	\hspace*{\algorithmicindent} \textbf{Input:} {The Polarity Truth Table  of the Sbox S, with $2^n$ Rows and $2^m-1$ Columns } \\
	\hspace*{\algorithmicindent} \textbf{Output:} {The FWHT of the Sbox S, with $2^n$ Rows and $2^m-1$ Columns   }
	\begin{algorithmic} 
		\STATE $z \leftarrow 0$
		\STATE $j \leftarrow 1$
		\STATE $WT \leftarrow PTT$
		\WHILE{$z < 2^m-1$}
		\WHILE{$j < 2^n$}
		\FOR{$i = 0$ to $2^n-1$}
		\IF{$i.j$$ = 0$} 
		\STATE $temp \leftarrow WT[i][z]$
		\STATE $WT[i][z] \leftarrow WT[i][z] + WT[i+j][z]$
		\STATE $WT[i+j][z]  \leftarrow temp$ $–$ $ WT[i+j][z]$
		\ENDIF
		\ENDFOR
		\STATE $j \leftarrow 2*j$
		\ENDWHILE
		\STATE{$z \leftarrow z+1$}
		\ENDWHILE
	\end{algorithmic}
\end{algorithm}

The nonlinearity of the S-Box S is the minimum nonlinearity of all linear combinations $\{g_j(x) \ , \ 0\le j \le 2^m-1\}$ (Equation \ref{nl}) of it's component functions\cite{17}.

\begin{equation} \label{nl}
\begin{split}
g_j\left(x\right) 
=\sum^{m-1}_{i=0}{a^{\left(j\right)}_if_i\left(x\right)} \ \ \ \
\left(0\le j\le 2^m-1,a^{\left(j\right)}_i\in F_2\right)
\end{split}
\end{equation}

The nonlinearity degree of the S-Box S ($nl(S)$) is the minimum Hamming distance of all functions of $v.S(x)$ with the Affine functions of $w.x$ and defined based on Equation \eqref{nl2} \cite{18}.

\begin{equation} \label{nl2}
\begin{split}
nl\left(S\right) 
=2^{n-1}-\frac{1}{2}{\mathop{\mathrm{max}}_{v\epsilon f^m_2,w\epsilon F^n_2} \left|\sum_{x\epsilon F^n_2}{{(-1)}^{<v,S\left(x\right)>\bigoplus <u,v>}}\right|\ }\ 
\left(0\le u\le 2^n-1\right),\left(0\le v\le 2^m-1\right)
\end{split}
\end{equation}

Al. \ref{algorithm2} shows the  nonlinearity pseudo-code \cite{14}.

\noindent 
\begin{algorithm}
 \small
	\caption{Pseudcode of Computing Nonlinearity of Sbox}
	\label{algorithm2}
	\hspace*{\algorithmicindent} \textbf{Input:} {The FWHT of the Sbox S, with $2^n$ Rows and $2^m-1$ Columns } \\
	\hspace*{\algorithmicindent} \textbf{Output:} {The Nonlinearity degree}
	\begin{algorithmic} 
		\STATE $temp \leftarrow 0$
		\STATE $Nonlinearity \leftarrow 10000$
		\STATE $var[2^m]  \leftarrow 0$
		\WHILE{$z < 2^m-1$}
		\WHILE{$j < 2^n$}
		\STATE \COMMENT{start Time-consuming part}
		\IF{$abs(WT[j][z]) > temp$} 
		\STATE $temp \leftarrow abs(WT[j][z]$
		\ENDIF
		\STATE $var[z] = ( 2^n - temp) >> 1$
		\STATE \COMMENT{end Time-consuming part}
		\ENDWHILE
		\IF{$var[z] < Nonlinearity$} 
		\STATE $Nonlinearity = var[z]$
		\ENDIF
		\STATE $j \leftarrow j+1$
		\STATE{$z \leftarrow z+1$}
		\ENDWHILE
	\end{algorithmic}
\end{algorithm}

\section{Definition of the Problem and Challenges} \label{dpc}

Evaluation of some important features of S-Boxes, such as nonlinearity, requires computation of walsh matrix. However, such calculations involve high levels of computational and memory complexity. The main purpose of this paper is to accelerate implementation of walsh matrix and nonlinearity computation for large S-Boxes.

The walsh matrix of large S-Boxes occupy a large amount of memory, and as the input and output of the S-Box increase, the number of accesses to its elements within the al.s increases exponentially. Consequently, computation of nonlinearity for large S-Boxes takes a long time. To accomplish this objective, several secondary objectives should be accomplish. Our secondary objectives include (a) improving parallelization of the FWHT, (b) using the walsh-hadamard matrix transpose, and (c) intelligently combining sub-problems (performing different operations with a single walsh matrix reading).

Furthermore  as described in \cite{9,11,13}, the previously proposed solutions for FWHT parallelization were limited to boolean functions and this paper attempt to extend it for parallelization of nonlinearty al. for large S-Boxes \cite{4,16}. In the next section, our proposed solutions for these challenges is proposed.

\section{Proposed Solutions} \label{ps}

At first we address (a)the issue of how to accelerate walsh matrix access using walsh matrix transpose. Then (b) we discuss reducing computational complexity of FWHT for S-Boxes and introduce our solution, which is a parallelized al. fitted to S-Boxes. Our third solution (c)is to reduce the computational and memory complexities in computing nonlinearity of a S-Box using a specific combination of it's sub-problems so that we perform nonlinearity computation for the S-Box (i.e. the time-consuming part) at the same time as walsh matrix computation. In the following, we describe these proposed solutions in the abovementioned order.

\subsection{Accelerating Access to Walsh Matrix using Walsh Matrix Transpose}

To change how the walsh matrix is defined and accessed, our solution is using the walsh matrix transpose instead of the walsh matrix itself in al.s \ref{algorithm1} and \ref{algorithm2}. For large matrices, since the compiler is not able to define large matrices, we need to define the data via manual allocation. Given the definition of the walsh matrix, this matrix (by default) must have $2^n$ rows of arrays with $2^m\mathrm{-}\mathrm{1}$ members (integers). Therefore, since the compiler needs to have access to all elements of each column of the matrix within the first loop of walsh al., it is necessary for the first loop of al. \eqref{algorithm1} to have access to all elements of each column and be able to read and modify them. At each round of the first loop of the walsh al., the array pointer shifts $2^n$ times (the number of walsh matrix rows) and the operating system brings the entire arrays or portions of them into the RAM. The timing of these operations, therefore, exponentially depends on the number of arrays exchanged between the RAM and the memory and pointers shifts. In other words, with the increase in the number of input and output bits of the S-Box, the number of first loop rounds and the number of arrays increase exponentially, resulting in much longer time-consuming pointer and array swap operations.

To solve this problem, the walsh matrix transpose is used instead of the walsh matrix itself. The transpose matrix is defined with $2^m\mathrm{-}\mathrm{1}$ rows of arrays with $2^n$ members (integers). Rows and columns are then replaced during the call in fast walsh al., simply instead of calling WT[i][z], we call WT[z][i]. As a result, in each round of the loop, only one array is called and by reducing pointer shifts and calling the array to a number, the array swap operations reduces effectively in each round of the first loop of al. \eqref{algorithm1}. For instance, for m=3 and n=4, as a result of using the walsh matrix itself at each round of the first loop of al. \eqref{algorithm1}, the operating system (depending on its structure) will bring the entire array into the RAM at least once and writes in the memory after the shift. Therefore, given the number of rounds in the first loop itself, at least 122 arrays will exchange between the memory and the RAM. This number reduces to 7 if we use the walsh matrix transpose because in each round of the first loop only one array will be brought into the RAM.

\subsection{Reducing Computational Complexity of Walsh Transform by Improving FWHT Parallelization}

Our second proposed solution is to reduce computational complexity of al. \eqref{algorithm1} using a parallelization method different from the previous methods in \cite{8} and \cite{9}. Previous methods were designed to parallelize the FWHT for a boolean function, which requires adding a new loop (to the iterations number of output bits) to calculate the FWHT of a S-Box. In this case, due to all the threads falling into one main loop, parallelization is not optimally performed and the operating system has to put threads into sleep modes after each round of the loop and recall them again. This requires a long exchange process.

\begin{figure}
	\includegraphics[width=0.6\linewidth]{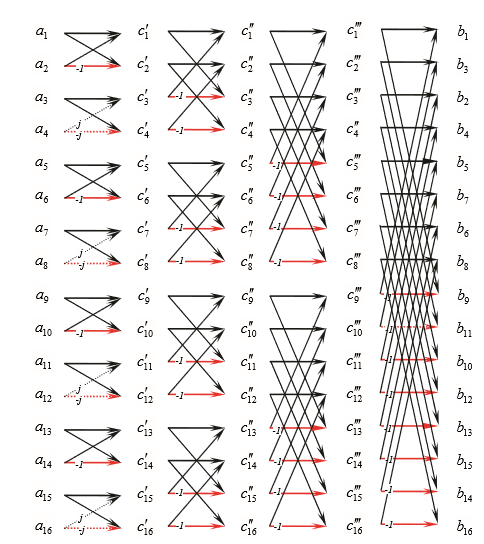}
	\centering
	\caption{The Fast Walsh Algorithm Process}
	\centering
	\label{fig1}
\end{figure}

Fig. \eqref{fig1} shows the process of al. \eqref{algorithm1} for just one column of the walsh matrix of a $4\times 4$ S-Box. According to al. \eqref{algorithm1}, calculations of walsh matrix columns are not dependent on each other, meaning that each column performs the operation of Fig. \eqref{fig1} independently. Therefore, all column calculations can be done concurrently and in parallel for parallelization.

The number of columns divides by the number of threads, that is, the number of rounds of the first loop in al. \eqref{algorithm1} divides by the number of threads that the processor can provide. The product of this operation represents the number of walsh matrix columns the calculation of which is assigned to each thread. For instance, computation of the walsh matrix with 8 thread for m=4 and n=3, the first loop of al. \eqref{algorithm1} consists of 15 rounds, each round assigned to calculate one column of the walsh matrix. Therefore, each thread is assigned to computation of two columns, save for the last one that will only be assigned to a single column. By doing so, we will achieve an acceleration of 7 folds on paper. The results of experiments can be found in Section 6. The walsh Parallelized al. pseudo-code (after making modifications to implement the next proposed solutions) is presented in al. \eqref{algorithm4}.

\subsection{Reducing Computational and Memory Complexity in Computation of Nonlinearity by Combining Sub-Problems}

Our third solution is to accelerate computation of nonlinearity of a S-Box by inserting the time-consuming part of the operation into the walsh matrix computation al. so that the compiler can perform the nonlinearity computation by a one-time call of the walsh matrix in the fast walsh computation al.. The al. for calculating nonlinearity of S-Boxes is presented in Section 3 as al. \eqref{algorithm2}. The fast walsh al. is changed into al. \eqref{algorithm3} by incorporation of the main part of nonlinearity computation al. (the time-consuming part in al. \eqref{algorithm2}) into al. \eqref{algorithm1} and using walsh matrix transpose instead of walsh matrix itself.
\noindent
\begin{algorithm}
    \small
	\caption{Pseudcode of FWT with Part of the calculations Nonlinearity}
	\label{algorithm3}
	\hspace*{\algorithmicindent} \textbf{Input:} {The Polarity Truth Table of the Sbox S, with $2^m-1$ Rows and $2^n$ Columns } \\
	\hspace*{\algorithmicindent} \textbf{Output:} {The Walsh Transform of the Sbox S, with $2^m-1$ Rows and $2^n$  Columns, The var array with $2^m$ entries }
	\begin{algorithmic} 
		\STATE $z \leftarrow 1$
		\STATE $j \leftarrow 1$
		\STATE $WT \leftarrow PTT$
		\STATE $temp \leftarrow 0$
		\WHILE{$z < 2^m-1$}
		\WHILE{$j < 2^n$}
		\FOR{$i = 0$ to $2^n-1$}
		\IF{$i.j$$ = 0$} 
		\STATE $temp \leftarrow WT[z][i]$
		\STATE $WT[z][i] \leftarrow WT[z][i] + WT[z][i+j]$
		\STATE $WT[z][i+j]  \leftarrow temp$ $–$ $ WT[z][i+j]$
		\IF{$i = 2^n-1$} 
		\IF{$abs(WT[z][j])$ $ > temp$} 
		\STATE $temp \leftarrow abs(WT[z][j])$
		\ENDIF
		\ENDIF
		\ENDIF
		\ENDFOR
		\STATE $j \leftarrow 2*j$
		\ENDWHILE
		\STATE{$var[z]$ $=  ( 2^n  - temp) >> 1$}
		\STATE{$z \leftarrow z+1$}
		\ENDWHILE
	\end{algorithmic}
\end{algorithm}
\noindent
\begin{algorithm}
 \small
	\caption{Pseudcode of Parallel Implementation of FWT}
	\label{algorithm4}
	\hspace*{\algorithmicindent} \textbf{Input:} {The Polarity Truth Table of the Sbox S, with $2^m-1$ Rows and $2^n$ Columns } \\
	\hspace*{\algorithmicindent} \textbf{Output:} {The Walsh Transform WT of the Sbox S, with $2^m-1$ Rows and $2^n$  Columns, The var array with $2^m$ entries }
	\begin{algorithmic} 
		\STATE $Create$ $FWT$ $threads$
		\STATE $Calling$ $FWT$ $threads$
		\STATE $Wait$ $for$ $FWT$ $threads$ $to$ $finish$
		\STATE $Joining$ $the$ $FWT$ $threads$
	\end{algorithmic}
\end{algorithm}
\noindent
\begin{algorithm}
    \small
	\caption{Pseudcode of threads FWT}
	\label{algorithm5}
	\hspace*{\algorithmicindent} \textbf{Input:} {The Polarity Truth Table of the Sbox S, with (($2^m-1$/Total number of threads) * number thread) Rows and $2^n$ Columns } \\
	\hspace*{\algorithmicindent} \textbf{Output:} {The Walsh Transform of the Sbox S, with (($2^m-1$/Total number of threads) * number thread) Rows and $2^n$  Columns, The var array with ($2^m$/ Total number of threads) entries }
	\begin{algorithmic} 
		\STATE $z \leftarrow 1$
		\STATE $j \leftarrow 1$
		\STATE $WT \leftarrow PTT$
		\STATE $temp \leftarrow 0$
		\STATE $waiting$ $for$ $call$ $by$ $main$ $function$
		\FOR{$index$ $from$ $((2^m-1$ $/$ $Total$ $number$ $of$ $threads)$ $*$ $(number$ $thread-1))$ $to$ $(( 2^m-1$ $/$ $Total$ $number$ $of$ $threads)$ $*$ $number$ $thread)$}
		\WHILE{$j < 2^n$}
		\FOR{$i = 0$ to $2^n-1$}
		\IF{$i.j$$ = 0$} 
		\STATE $temp \leftarrow WT[z][i]$
		\STATE $WT[z][i] \leftarrow WT[z][i] + WT[z][i+j]$
		\STATE $WT[z][i+j]  \leftarrow temp$ $–$ $ WT[z][i+j]$
		\IF{$i = 2^n-1$} 
		\IF{$abs(WT[z][j])$ $ > temp$} 
		\STATE $temp \leftarrow abs(WT[z][j])$
		\ENDIF
		\ENDIF
		\ENDIF
		\ENDFOR
		\STATE $j \leftarrow 2*j$
		\ENDWHILE
		\STATE{$var[z]$ $=  ( 2^n  - temp) >> 1$}
		\STATE{$z \leftarrow z+1$}
		\ENDFOR
		\STATE $the$ $main$ $function$ $of$ $completion$ $notifications$
	\end{algorithmic}
\end{algorithm}

Al. \eqref{algorithm3}, which is implementation of the fast walsh al. after applying all the acceleration methods, is parallelized in al. \eqref{algorithm4}.
Al. \eqref{algorithm4} calls threads that are designed and functioned in al. \eqref{algorithm5} and waits for all of them to finish. Inputs and outputs of threads are the same as inputs and outputs of the fast walsh al., except that each thread is responsible for computation of a column in the walsh matrix.

The number of columns and their span depends on the index and the total number of threads, which is clearly specified in al. \eqref{algorithm5}. This al. initially awaits to call the thread through walsh main function. Eventually, each thread informs the main function of completing its operation.
\noindent
\begin{algorithm}
\small
	\caption{Pseudcode of Computing Nonlinearity }
	\label{algorithm6}
     \hspace*{\algorithmicindent}\textbf{Input:} {The var array with $2^m$ entries } \\
	 \hspace*{\algorithmicindent}\textbf{Output:} {The Nonlinearity degree}
	\begin{algorithmic} 
		\STATE $temp \leftarrow 0$
		\STATE $Nonlinearity \leftarrow 10000$
		\STATE $var[2^m]  \leftarrow 0$
		\WHILE{$z < 2^m-1$}
		\IF{$var[z] < Nonlinearity$}
		\STATE $Nonlinearity = var[z] $
		\ENDIF
		\STATE{$z \leftarrow z+1$}
		\ENDWHILE
	\end{algorithmic}
\end{algorithm}

The nonlinearity computing al. \eqref{algorithm2} was changed to al. \eqref{algorithm6}. The initial calculation in al. \eqref{algorithm6}, which comprises the time-consuming part of al. \eqref{algorithm2}, is performed through al. \eqref{algorithm3} or its parallelized version, al. \eqref{algorithm4}. This al. captures an array containing the computation result for each of the walsh matrix columns (obtained from the fast walsh al. proposed in this paper) and returns the nonlinear degree as the output by finding the smallest number in the array. 
The resource consumption of proposed and previous al.s as well as time and memory complexities of each walsh al. and the nonlinearity computation are compared in Table \eqref{tb1}. In calculating the complexities, it is assumed that the S-Box is $n\times m$. In each round of the loop in nonlinearity compotation al., we have access to FWHT of S-Boxes that have been calculated before and are accessible in each memory and thus, we do not need to recalculate this transform. Therefore, the FWHT does not affect the time order of the nonlinearity computations.
\noindent
\begin{table}[!t]
    \centering \small
	\caption{Comparison of the Computational and Memory Complexity of algorithms}
	\centering
	\label{tb1}
	\centering
	\begin{tabular}{|p{1.3in}|p{0.5in}|p{0.8in}|p{1.0in}|p{0.8in}|} 
		\hline 
		Complexity Order \newline  & FWHT \newline al. \eqref{algorithm1} & Proposed FWHT\newline al. \eqref{algorithm4}  & Nonlinearity Computation of \newline al. \eqref{algorithm2} & Proposed Nonlinearity of al. \eqref{algorithm6} \\
		\noalign{\smallskip} \hline 
		Computation (time) & ${n2}^n2^m$ & ${n2}^n{(2}^m/TNT)$ & $2^n2^m$ & $2^n$ \\ \hline 
		Memory & $2^n2^m$ & $2^n2^m+2^n$ & $2^n2^m$ & $2^n$ \\ \hline 
		\multicolumn{5}{|p{4in}|}{Total\_number\_threads:TNT, The Integers are 8 bits.} \\ \hline 
	\end{tabular}
 	\end{table}

According to Table \eqref{tb1}, the time and memory complexities of fast walsh al.s are exponential. For the time complexity, the fast walsh computation al. consists of two nested loops, and the memory complexity corresponds with the memory of walsh matrix memory, which is ${O(2}^{n+m})$  for S-Boxes. Within our proposed fast walsh, we have a walsh matrix and an array for basic nonlinearity computations. Therefore, the memory complexity is equal to the sum of memory they occupy.

\section{Results of Empirical Experiments} \label{ree}

To ensure the validity of the proposed al.s, the nonlinearity of the AES S-Box and two other S-Boxes is compared with both the initial al. \eqref{algorithm1} and the proposed one al. \eqref{algorithm6} (including the combination of all methods). As shown in Table \eqref{tb2}, the results of both al.s were perfectly consistent.
\noindent
\begin{table}[!t]
	\centering
	\caption{Comparison of Nonlinearity Computations}
	\centering
	\label{tb2}
	\centering
	\begin{tabular}{|p{1.5in}|p{0.6in}|p{0.6in}|p{0.6in}|} \hline 
		Size of S-Box & 8$\mathrm{\times}$8 & 10$\mathrm{\times}$10 & 16$\mathrm{\times}$16 \\ \hline 
		Initial al.- al. \eqref{algorithm1} &112& 384 & 32,000 \\ \hline 
		Proposed al.- al. \eqref{algorithm6} &112& 384 & 32,000 \\ \hline 
	\end{tabular}
\end{table}

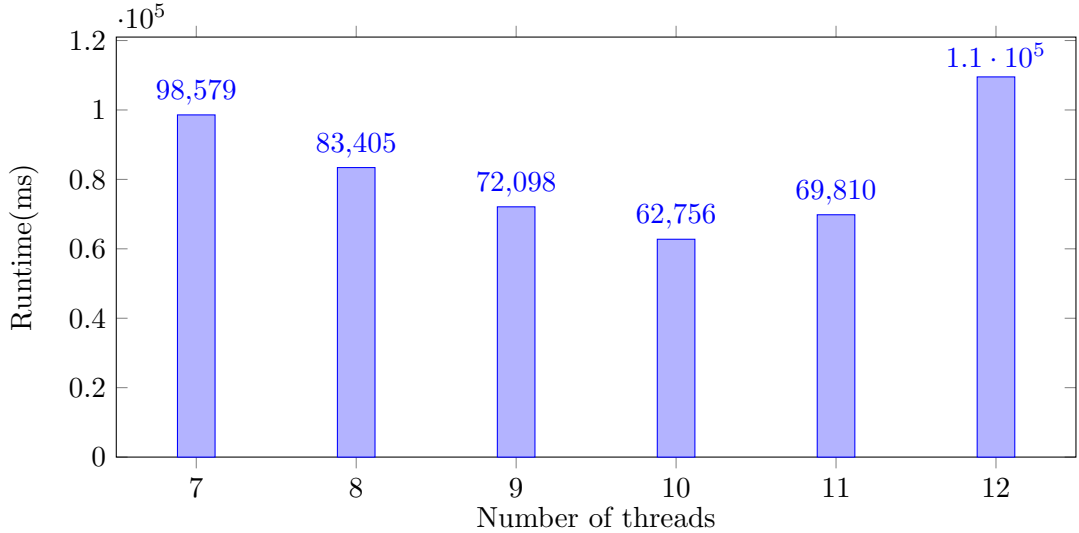
\begin{figure}    
	\centering
	\begin{tikzpicture}
	\begin{axis}[
	ybar,
	bar width=.5cm,
	width=\textwidth,
	height=.5\textwidth,
	legend style={at={(0.46,1)},
		anchor=north,legend columns=0},
	symbolic x coords={7,8,9,10,11,12},
	xtick=data,
	nodes near coords,
	nodes near coords align={vertical},
	ymin=0,ymax=121000,
	ylabel={Runtime(ms)},
	xlabel={Number of threads},
	]
	\addplot table[x=interval,y=carT]{\mydataa};
	\end{axis}
	
	\end{tikzpicture}
	\caption{Run the al. \eqref{algorithm4} on the number of different threads}
	\label{ch1}
\end{figure}

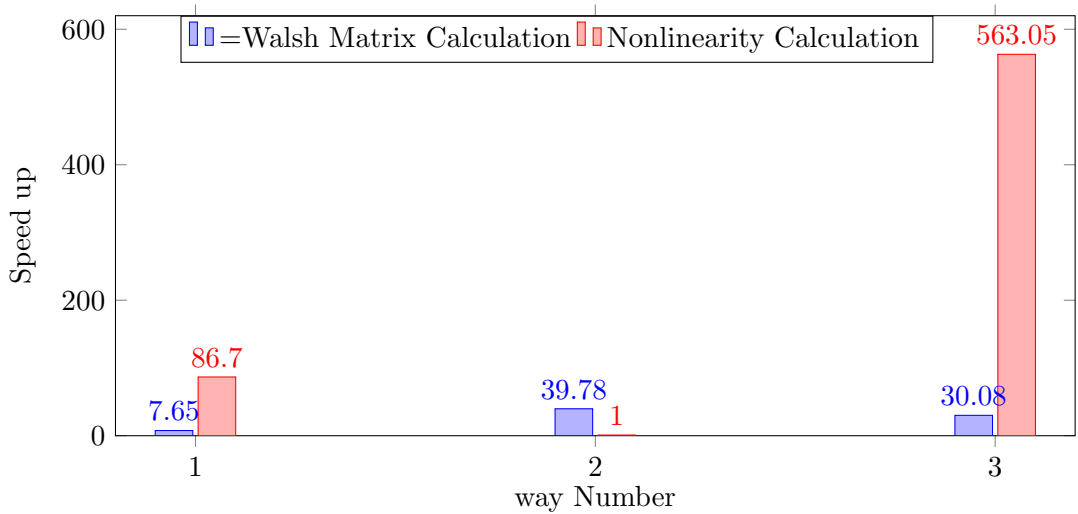
\begin{figure}    
	\centering
	\begin{tikzpicture}
	\begin{axis}[
	ybar,
	bar width=.5cm,
	width=\textwidth,
	height=.5\textwidth,
	legend style={at={(0.46,1)},
		anchor=north,legend columns=0},
	symbolic x coords={1,2,3},
	xtick=data,
	nodes near coords,
	nodes near coords align={vertical},
	ymin=0,ymax=620,
	ylabel={Speed up},
	xlabel={way Number},
	]
	\addplot table[x=interval,y=carT]{\mydata};
	\addplot table[x=interval,y=carD]{\mydata};
	\legend{=Walsh Matrix Calculation,Nonlinearity Calculation}
	\end{axis}
	\end{tikzpicture}
	\caption{The speed up of each of the methods for S-boxs 16$\mathrm{\times}$16}
	\label{ch2}
\end{figure}

Three S-Boxes of different sizes (two larger boxes were produces with Sage software) were used in comparison of time complexity. Each of the numerical results obtained (Figs. \eqref{ch1} and \eqref{ch2}) is the average of several iterations of al.s. Table \eqref{tb3} presents specifications of the computational system on which the evaluations are performed.
\noindent
\begin{table}[!t]
	\centering
	\caption{Specifications of Process System}
	\centering
	\label{tb3}
	\centering
	\begin{tabular}{|p{1.0in}|p{2.2in}|} \hline 
	Environment&	Platform   \\ 
		\hline 
CPU	&	Inter Core i7 8400K-12 thread   \\ 
		\hline 
 Memory(RAM)	&	16Gb DDR4  \\ 
		\hline 
 OS	&	Win10 x64 \\ 
		\hline 
Compiler	&	MSVC 2017   \\ 
		\hline 
	\end{tabular}
\end{table}

First, the fast walsh al. is implemented in parallel on different number of threads repeatedly. The average runtime results for different number of threads are shown in Fig. \eqref{ch1}. This diagram shows that 10-thread implementation has the largest runtime reduction. This seems to be due to the fact that the level of parallelization decreases with fewer threads. In the cases with more than 10 threads, the operating system is forced to put some of the threads into sleep mode and wake them again in order to run other threads; this will increase runtime.

\pgfplotstableread[row sep=\\,col sep=&]{
	interval & carT  \\
	7     & 98579   \\
	8     & 83405 \\
	9    & 72098 \\
	10     & 62756   \\
	11    & 69810 \\
	12     & 109525   \\
}\mydataa
\noindent

Table \eqref{tb4} indicates the runtime of the fast walsh al. and nonlinearity computation for S-Boxes of 10 $\mathrm{\times}$ 10, 12 $\mathrm{\times}$ 12 and 16 $\mathrm{\times}$ 16 in milliseconds for the initial methods and proposed ones. Fig. \eqref{ch2} presents the runtime reduction rate for each of the methods in Table \eqref{tb4}, which is the ratio of the runtime of the primary al. (method) to the runtime of the proposed al.s (methods) for a 16 $\mathrm{\times}$ 16 Box. This number is always greater than one, indicating that proposed methods 1, 2, and 3 are faster than the primary methods.

\noindent
\begin{table}[!t]
	\centering 
	\caption{Runtime of Proposed Methods in Milliseconds (m.s.)}
	\label{tb4}
	\begin{tabular}{|p{1.3in}|p{0.6in}|p{0.8in}|p{0.7in}|p{0.5in}|p{0.5in}|} \hline 
	 Method (m.s.)& Size& Primary al.	 &  (a)&  (b) &   (c)    \\ 
		\hline 
	 FWHT 	&\centering 10 $\mathrm{\times}$ 10& 160  & 48 & 17& 20 \\ 
		\hline 
		FWHT 	& \centering 12 $\mathrm{\times}$ 12& 1641&452 & 98 & 120\\ \hline 
	FWHT 	&\centering 16 $\mathrm{\times}$ 16	& 2,496,726   & 8,326,331 & 62,756&83,013   \\ 
		\hline 
	Nl. Computation 	&\centering 10 $\mathrm{\times}$ 10  & 372 & 7 & 7 & 20  \\ 
		\hline 
	Nl. Computation 	&\centering 12 $\mathrm{\times}$ 12   	&2736 & 38 & 38  & 120  \\ \hline 
	Nl. Computation 	&\centering 16 $\mathrm{\times}$ 16 & 46,741,627    & 539,120 & 539,120	&83,015   \\ \hline 
	\end{tabular}
\end{table}

\pgfplotstableread[row sep=\\,col sep=&]{
	interval & carT & carD \\
	1     & 7.65  & 86.699 \\
	2     & 39.784 & 1  \\
	3    & 30.076& 563.05 \\
}\mydata

As shown in Fig. \eqref{ch2} and Table \eqref{tb4}, the number of displacements between memory and pointer shifts has reduced exponentially by the first proposed method (using the walsh matrix transpose instead of the walsh matrix).

\noindent 

The walsh al. runtime for a 16 $\mathrm{\times}$ 16 S-Box has reduced from 2,496,726 milliseconds to 326,331 milliseconds (runtime improvement of 7.65 times). The runtime has also increased to 62.75 milliseconds after parallelization with 10 threads (which is an improvement of 5.19 times compared to the first method). While we expected to achieve an improvement of 10 folds after parallelization with 10 threads, we failed to meet the expectation because the operating system does not run the threads completely concurrently and the arrays pointer shifts and the arrays swaps concurrently imposes time load. However, it can be seen that we achieved a runtime improvement of 39.784 times in computation of FWHT by parallelization and using the walsh matrix transpose.

Moreover, by employing the modified al. \eqref{algorithm4} the combination of proposed acceleration methods of 1 and 2) in nonlinearity computation for a 16 $\mathrm{\times}$ 16 S-Box, the runtime of al. \eqref{algorithm1} reduced from 46,741,627 milliseconds to 539,120 milliseconds. This marks an 86.69-fold improvement. This is due to a decrease walsh matrix elements' swap and the pointer shifts. We continues by incorporating part of the computation process into parallelized fast walsh al. \eqref{algorithm6} (a combination of all the proposed acceleration methods) to reduce the runtime to 83,015 milliseconds and reach an improvement of 6.49-fold compared to the primary al..

\section{Conclusion} \label{co}

The main purpose of this paper was to decrease the runtime of FWHT al. and nonlinearity computation for large S-Boxes. To meet this goal, three methods were proposed. Based on Table \eqref{tb4}, our best choice for FWHT computation resulted in a runtime improvement of 39.784-fold and for nonlinearity computation using walsh matrix transpose a runtime improvement of 563.05-fold. 

As explained in the previous sections, nonlinearity computation of S-Boxes can be concluded in much lower time using methods proposed in this paper and as a result, more S-Boxes can be examined in less time.

there exist two suggestions for future research. First, to implement similar methods for computing other features of S-Boxes such as balancedness, propagation criterion, bit independence, and correlation immunity using FWHT. Second, given the immense possibilities for parallelization on GPUs, to implement proposed solutions of this paper on GPUs to evaluate S-Boxes

\end{document}